# Our House is Our Glassy Castle: Challenges of Pervasive Computing in Private Spaces

**Dinislam Abdulgalimov**
Open Lab, Newcastle University,
Newcastle upon Tyne, UK
d.abdulgalimov@ncl.ac.uk

**Timur Osadchiy**
Open Lab, Newcastle University,
Newcastle upon Tyne, UK
t.osadchiy@ncl.ac.uk

## Abstract
Modern society is going through the transformation under the influence of the Information Technologies. Internet of Things as one of the latest facet of it becoming more visible and widely spread. We wish to reflect and discuss the current concerns regarding its' expansion. Our particular interests lie in the increasing of the usability and comfortability through the unification of the IoT protocols and security measures. As well as addressing the privacy concerns and discussing of the possible changings in the perception of the privacy and personal space concepts.

## ACM Classification Keywords
H.5.m. Information interfaces and presentation (e.g., HCI): Miscellaneous.

## IoT in everyday life
Today the "Internet of things" (IoT) are not just buzz words that draw the interest of the investors and the big companies but rather the fast-developing field that represents a major departure in the history of the Internet. IoT moves it beyond the rectangular scopes of the desktop computers, tablets and smartphones and helps to power up millions of the everyday devices from the kettles, home thermostats and light bulbs. According to Business Insider the predicted amount of installed IoT devices will exited 20 billion by 2020 [1], based on the current Global World Wide Web penetration rate [2] and the changes of the hardware prices [3]. The Internet of Things is expected to have significant impact on many, if not all, aspects of our daily living such as public, work or private spaces. IoT aims to change the approach of individual existence in the environment that is "seamlessly equipped" with technologies that enable it to be sensed and controlled remotely through digital networks using computer-based systems, improving their efficiency and providing a variety of economic and social benefits. The future that advertised to us by IoT devices' vendors appears to be unshadowed and clear [4][5]. Is it?

**Unification and Security concerns**

As any rapidly changed and popular technology or field of human activity, IoT is currently going through the stage of formation, and acceptance by the general public and business. These processes attract the attention of the variety of researches and organizations with their own methods and views on the way how things need to be delivered and deployed [8][9][10][11]. The overall situation is similar to the beginning of the Internet era, when each of the major players tried to introduce and enforce their own vision before they managed to agree on the unified standards and approaches. In addition to that, because of the increased interest to the field, companies strive to introduce their solutions, products and specific innovations with the minimum time frame, while there are still unexplored domains and aspects of the human life where smart-devices can be used. And again, similarly to the situation with the early days of the World Wild Web, modern IoT products mainly concentrate on the delivery of the intendent service giving a low priority to security, reliability and resilience [6][12]. As a result, this creates a heterogeneous and chaotic situation in the IoT field.

What does this mean for us – general users?

- Uncertainty regarding the presence or absence of the vulnerabilities in the specific system or device we want to use;
- Inability to receive a seamless and "natural" experience in case of using similar products from different vendors, due to absence of unified protocol;
- Vendor lock. In case we want to get the comprehensive experience, we could achieve it only if a vendor has complete product line.

Worth to mentioned that there are Open Standards that aim to unite all the devices or provide the unified API for third party systems and devices [10]. However, they are not widely accepted by manufacturers.

We are keen to examine and discuss the aspects of aforementioned challenges, their impact on the user perception of smart houses, as well as usability and safety/security issues that can affect the trust towards IoT and potential ways to overcome those challenges.

**Privacy concerns**

Another concern that directly linked with the issue of vanishing agency of the individual is the privacy in the modern field of the IoT. It implies that the interconnected everyday objects (IoT devices/things) and environments that are "seamlessly equipped" with technologies are connected to the global (semi-global – tied to one vendor) network that one way or another can receive the data from the IoT devices. Partially because of the impossibility of delivering the same quality service without using powerful remote backend (servers' infrastructure) and partially because of the demand from the big companies for more personal and specific information about users. Thus, by introducing smart devices and systems to our everyday activity we willingly provide third parties with the big amount of personal information [13] [14]. It is even better than the famous postulate: "If you're not paying for it, you become the product" [17].

Those of us who consider our home or car to be primarily personal spaces where we can escape from the outside world (even for a short amount of time), would be reluctant to accept these rules. Additionally, as IoT developers try to attract and engage third party developers into using their platforms and services they tend to open access to their services or provide access to "anonymised" data for research purposes [15]. Previous researches showed that in some cases aggregation of even anonymised information still could lead identification of a person or the disclosure of one's information [16].

We wish to query the ways in which we can control information disclosure and manage the exchange process in more privacy-friendly way.

**Perception changing**

Our perception of the privacy and the personal space was coined through out of the human history. During the evolution of our society and the historical changes the perception changed from the Roman times to the 20$^{th}$ century [6]. In which the modern concept of these terms has been crystallized.

What if it's time for perception to change? Like it did before, affected by historical events and movements. What if the society with the overall process of the emerging technologies (such as IoT, Social Networks and the Internet itself) stands at the origins of the new evolutional step? What if the privacy and agency as we knew them before are outdated today? What if it is time to thing and explore the new notion? Maybe the concepts of the private space are not relevant anymore. The fact that IoT devices somehow collect, process and possibly make it accessible to others is just a step to the rethinking of the norms and concepts. A general user hypothetically law-abiding and doesn't have anything to hide. Moreover, in case of the emergency (accident, health problem) IoT, if monitoring, could request emergency services or other type of help.

Maybe we need to reconsider our beliefs? Maybe it's time for the brave new world?

This particular point is interesting one and in our opinion is worth to discuss in more depth during the workshop. All smart devices and sensors around us collecting a lot of information and sending it to different companies and organisation. It could be portrayed as a more efficient way of preserving safety and security. You as a user always know that there are some big companies/organisation that watching and analysing your habits and actions with potential to make your life easier and more efficient as well as safer. If only you'll overcome this ancient need to preserve privacy and personal space.

**Biographies**
**Dinislam Abdulgalimov** is a Doctoral Trainee in Digital Civics at Open Lab, Newcastle University, UK. His academic background was in the field of information security. He is interested in community collaboration and the possibility of using crowd-sourcing techniques for gathering operative information, exploring privacy versus trust aspects in such systems and examining the possibility of using peer-to-peer and decentralised models without exposing identity.

**Timur Osadchiy** is a Doctoral Trainee at the Open Lab, Newcastle University, UK. His academic background was in the field of computer science and applied cyber security and developing user-friendly communication protocols within IoT aiming in making it seamless for people with no technical background.